\documentclass[aps,prb,reprint,superscriptaddress,floatfix, nolongbibliography]{revtex4-2}

\usepackage{amsmath,amsfonts,amssymb}
\usepackage{dcolumn}
\usepackage{graphicx}
\usepackage[utf8]{inputenc}
\usepackage[english]{babel}
\usepackage{multirow}
\usepackage[colorlinks=true, allcolors=blue]{hyperref}

\graphicspath{{../figures/}}

\newcommand{\figwidthone}{0.8}  
\newcommand{\figwidthtwo}{0.8}  
\newcommand{\figwidththree}{0.9}
\newcommand{\figwidthfour}{0.6}

\begin{document}

\title{Electron-Nuclear Interactions as a Test of Crystal-Field Parameters for Low Symmetry Systems: Zeeman-Hyperfine Spectroscopy of Ho$^{3+}$ Doped {Y$_2$SiO$_5$}}

\author{Sagar Mothkuri}
\affiliation{School of Physical and Chemical Sciences, University of Canterbury, PB 4800, Christchurch 8140, New Zealand}
\affiliation{The Dodd-Walls Centre for Photonic and Quantum Technologies, New Zealand}
\author{Michael F. Reid}
\email[Corresponding author: ]{mike.reid@canterbury.ac.nz}
\author{Jon-Paul R. Wells}
\email[Corresponding author: ]{jon-paul.wells@canterbury.ac.nz}
\affiliation{School of Physical and Chemical Sciences, University of Canterbury, PB 4800, Christchurch 8140, New Zealand}
\affiliation{The Dodd-Walls Centre for Photonic and Quantum Technologies, New Zealand}
\author{Elo{\"i}se Lafitte-Houssat}
\affiliation{Chimie ParisTech, PSL University, CNRS, Institut de Recherche de Chimie
Paris, 75005 Paris, France}
\affiliation{Thales Research and Technology, 1 Avenue Augustin Fresnel, 91767, Palaiseau, France}
\author{Philippe Goldner}
\affiliation{Chimie ParisTech, PSL University, CNRS, Institut de Recherche de Chimie
Paris, 75005 Paris, France}
\author{Alban Ferrier}
\affiliation{Chimie ParisTech, PSL University, CNRS, Institut de Recherche de Chimie
Paris, 75005 Paris, France}
\affiliation{Facult{\'e} des Sciences et Ing{\'e}nierie,  Sorbonne Universit{\'e}, UFR 933, 75005 Paris, France}

\date{\today} 

\begin{abstract}
High-resolution Zeeman spectroscopy of electronic-nuclear hyperfine levels of  ${^5\mathrm{I}_8}\rightarrow{^5\mathrm{I}_7}$ transitions in Ho$^{3+}$:Y$_2$SiO$_5$ is reported. Crystal-field parameters determined for the two $C_1$  symmetry sites in Er$^{3+}$:Y$_2$SiO$_5$ are successfully used to model the Zeeman-hyperfine data, including the prediction of avoided crossings between hyperfine levels under the influence of an external magnetic field. The two six- and seven-coordinate substitutional sites may be distinguished by comparing the spectra with crystal-field calculations.
\end{abstract}
\maketitle

\section{Introduction}

Yttrium orthosilicate (Y$_2$SiO$_5$) doped with rare-earth ions is widely seen as an attractive option for the development of quantum-information technologies. The magnetic moment of yttrium  is very  small and  isotopes of Si and O with non-zero nuclear spin have very low natural abundances.  Consequently, decoherence due to spin flips is low, giving outstanding coherence properties. Furthermore, the substitutional sites for rare-earth ions in Y$_2$SiO$_5$ have $C_1$ point-group symmetry, giving  highly admixed wavefunctions, which enables efficient and diverse optical pumping schemes
\cite{rippe2008,lauritzen2008,longdell_experimental_2004}.  
Applications  include optical quantum memories \cite{deRiedmatten2008,zhong_optically_2015,zhong2017,rancic2018,lauritzen2010,fraval_dynamic_2005},  quantum-gate implementations \cite{longdell_experimental_2004,rippe2008}, microwave-to-optical photon modulators \cite{probst2013,xavi2015},  single-photon sources \cite{thompson2018}. Recently, control of multiple ions at the single-photon level has been demonstrated \cite{ChMoPhOuTh2020}. 

Performance improvements for such technologies require highly accurate  modeling of
magnetic-hyperfine structure to determine optimized regions where the Zero-First-Order-Zeeman (ZEFOZ)
technique can be most efficiently exploited. Use of the ZEFOZ technique has already enabled the   demonstration of a spin coherence time of
six hours in $^{151}$Eu$^{3+}$:Y$_2$SiO$_5$ \cite{zhong_optically_2015}. 

Crystal-field calculations
\cite{carnall_systematic_1989,GoBi96,NeNg00,liu_electronic_2006} may be used to model
the electronic structure of the entire $4f^N$ configuration of a rare-earth ion. Since the crystal-field
parameters vary in a systematic way across the rare-earth series, the information they
carry may be transferred between different ions. However, the determination of crystal-field parameters for crystals where the rare-earth ions occupy 
low point-group symmetry sites is
non-trivial. Additional information from magnetic splittings is essential to provide the orientation
information necessary to determine a unique set of parameters
\cite{antipin1972,smcaf22018}.  In $C_1$ symmetry (i.e.\ no symmetry)
there are 27 crystal-field parameters, so the calculations are
computationally challenging. However, recently we have developed techniques that make it practical to obtain full phenomenological
crystal-field fits for $C_1$ point-group symmetry sites. These
methods have been applied to both sites of Er$^{3+}$:Y$_2$SiO$_5$
\cite{Horvath2016,Horvath2019} with considerable success.  These parameters have also been  applied to the Zeeman splittings of Nd$^{3+}$ and Sm$^{3+}$ ions in Y$_2$SiO$_5$  \cite{Jobbitt2019}, demonstrating that the parameters may be transferred from ion to ion, and that the two six- and seven-coordinate substitutional sites may be distinguished by comparing calculations with experiment.  

The Nd$^{3+}$ and Sm$^{3+}$ work did not include nuclear hyperfine measurements. In this work, we utilize the large magnetic moment and high nuclear spin ($I=7/2$) of the trivalent holmium ion, and the application of magnetic fields, to explore the predictive ability of crystal-field calculations in the context of Zeeman-hyperfine measurements. This analysis is directly relevant to the utilization of the ZEFOZ technique in quantum information storage.

\section{Experimental}

Y$_2$SiO$_5$ (in the X2 phase) is a monoclinic crystal with $C^6_{2h}$
space group symmetry. The yttrium ions occupy two crystallographically
distinct sites, each with $C_1$ point-group symmetry, with oxygen coordination numbers of six
and seven \cite{maksimov1971crystal}. We follow the labeling convention used in Er$^{3+}$ \cite{sun_magnetic_2008,Horvath2019} and in Ref.\  \cite{Jobbitt2019} of referring to these as Site 1 and Site 2, tentatively identified as six- and seven-coordinate. 
Y$_2$SiO$_5$ has
three perpendicular optical-extinction axes: the crystallographic $b$
axis, and two mutually perpendicular axes labeled $D_1$ and $D_2$. In
our calculations we follow the convention of identifying these as the
$z$, $x$, and $y$ axes respectively \cite{sun_magnetic_2008}.

The crystal of Ho$^{3+}$:Y$_{2}$SiO$_{5}$  used in the current study was prepared using the Czochralski process with a Ho$^{3+}$ concentration of ${200}$\,ppm. The crystal was oriented using Laue backscattering.  The sample was a cuboid with the $D_1$ and $D_2$ and $b$ axes through the faces and dimensions of approximately 5\,mm on each side. The sample was given a spectroscopic quality polish exhibiting excellent optical quality. 

Infrared absorption spectroscopy was performed using a Bruker
Vertex 80 FTIR having its entire optical path purged by N$_2$ gas. This instrument has a maximum apodized resolution of 0.075\,cm$^{-1}$ (2.2\,GHz). Zeeman measurements were performed using a 4\,T, superconducting, simple solenoid built into a home-built cryostat. The sample is mounted on a screw fitted into the bore of the solenoid and is therefore cooled by thermal contact with the 4.2\,K helium bath of the solenoid.

\section{Theoretical background}

 The Hamiltonian for the $4f^N$ configuration may be written as
\cite{carnall_systematic_1989,GoBi96,liu_electronic_2006}
\begin{equation}
  H = H_{\mathrm{FI}} + H_{\mathrm{CF}}  + H_{\mathrm{HF}} + H_{\mathrm{Z}} .
  \label{eqn:hdefn}
\end{equation} 
The terms in this equation represent
the free-ion contribution, the crystal-field interaction, the  electron-nuclear hyperfine interaction, and the Zeeman interaction. 

The free-ion Hamiltonian  may be written as
\begin{align}
 H_{\mathrm{FI}} &= E_\text{avg} + \sum_{k=2,4,6} F^k f_k + \zeta   
   \sum_i \mathbf{s}_i \cdot \mathbf{l}_i
 \nonumber\\
 &+\alpha L(L+1) + \beta G(G_2) + \gamma G(R_7) 
                 + \sum_{k = 2,3,4,6,7,8} T^k t_k
 \nonumber \\  
 &  + \sum_{k=0,2,4} M^k m_k + \sum_{k=2,4,6} P^k p_k . 
  \label{eqn:h_fi_defn}
\end{align}
$E_\text{avg}$ is a constant configurational shift, $F^k$ Slater
parameters characterizing aspherical electrostatic repulsion, and
$\zeta$ the spin-orbit coupling constant. The sum in the spin-orbit term is over the $4f$ electrons. The other terms 
parametrize two- and three- body interactions, as well as higher-order
spin-dependent effects
\cite{carnall_systematic_1989,liu_electronic_2006}.

The crystal-field Hamiltonian has the form 
\begin{equation}
    H_{\mathrm{CF}} = \sum_{k,q} B^{k}_q C^{(k)}_q , 
\end{equation}
with $k = 2$, $4$, $6$, $q = -k,\ldots,k$. The $B^{k}_q$ are crystal field parameters, $C^{(k)}_q$ are spherical tensor operators. In $C_1$ symmetry, all non-axial ($q \neq 0$)\ $B^{k}_q$ parameters are complex, leading to a total of 27 parameters. Due to this low symmetry, all electronic states are non-degenerate in Ho$^{3+}$.  

The holmium nucleus has a spin $I=7/2$, and the electronic states are coupled to the nuclear spins by the hyperfine interaction, giving 8 electronic-nuclear states for each electronic level. We use  basis states $|JM_J,IM_I\rangle$ in our calculations. 

For holmium the magnetic-hyperfine interaction is much larger than the nuclear-quadrupole interaction \cite{Wells2004} so here we consider only the magnetic hyperfine part of the Hamiltonian, which within a single multiplet may be written as
\begin{equation}
    H_{\mathrm{HF}} = A_{J}\mathbf{J}\cdot \mathbf{I}  , 
  \label{eqn:AdotI}
\end{equation}
where $A_J$ is the hyperfine coupling constant for the multiplet. The  $A_J$ may be calculated using eigenvectors of the free-ion part of the Hamiltonian. 

The effect of an external magnetic field is given by the Zeeman Hamiltonian
\begin{equation}
    H_{\mathrm{Z}} = \mu_B\mathbf{B} \cdot (\mathbf{L} +\mathbf{2S}) , 
  \label{eqn:Zeeman}
\end{equation}
where B is the  magnetic field. Within a multiplet $\mathbf{L} +\mathbf{2S}$ may be written as a $g$ factor $g(^{2S+1}L_J)$ multiplied by $\mathbf{J}$.

\begin{figure}[tb!]
\centering
\includegraphics[width=\figwidthtwo\columnwidth]{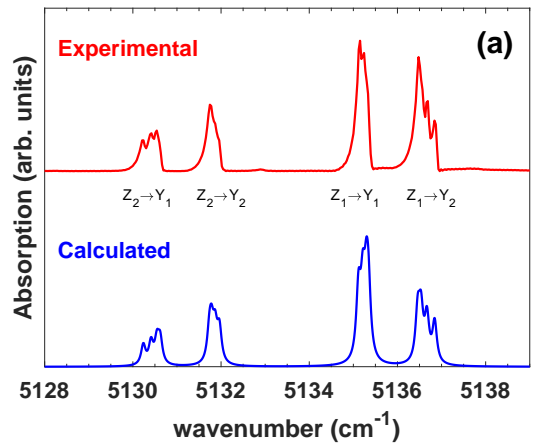}\\
\bigskip
\includegraphics[width=\figwidthtwo\columnwidth]{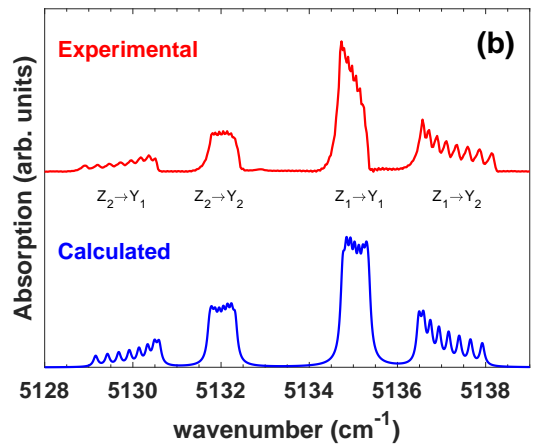}
\caption{\label{fig:Site_2_0_0.3T_Zeeman}  
4.2~K absorption spectrum of the Z$_{1}$-Z$_{2}\longrightarrow$Y$_{1}$-Y$_{2}$ transitions for Site 2 of Ho$^{3+}$:Y$_{2}$SiO$_{5}$. (a) Zero field spectrum and (b) 0.3~T spectrum with a magnetic field applied along the $b$ axis.}
\end{figure}

\section{Results and discussion} 

\begin{table}[tb!]
    \centering
    \caption{\label{tab:Site_2_0T_Para}
Experimental and calculated Z$_{1}$-Z$_{2}\longrightarrow$Y$_{1}$-Y$_{2}$ transition energies for Site 2. All values are in units of cm$^{-1}$.}
    \begin{tabular}{c c c}
    \hline
    \hline
Transition & Experimental & Calculated\\
\hline
Z$_{2}$-Y$_{1}$     & 5130.22 & 5130.23\\
                    & 5130.41 & 5130.41\\
                    & 5130.54 & 5130.54\\
                    & 5130.61 & 5130.62\\
\\
Z$_{2}$-Y$_{2}$     & 5131.68 & 5131.73\\
                    & 5131.76 & 5131.78\\
                    & 5131.87 & 5131.87\\
                    & 5131.95 & 5131.96\\
\\
Z$_{1}$-Y$_{1}$     & 5135.09 & 5135.11 \\
                    & 5135.14 & 5135.21\\
                    & 5135.24 & 5135.29\\
                    & 5135.32 & 5135.34\\
\\
Z$_{1}$-Y$_{2}$     & 5136.46 & 5136.45 \\
                    & 5136.49 & 5136.53\\
                    & 5136.68 & 5136.66\\
                    & 5136.84 & 5136.84\\
\hline
\hline
    \end{tabular}
\end{table}

\begin{table}[tb!]
\centering
\caption{\label{tab:Fitparameters}
Energy levels for the Z$_{1}$, Z$_{2}$, Y$_{1}$ and Y$_{2}$ states for Site 1 and Site 2 from crystal-field calculations (CF) and after fitting to the experimental data. The last row shows the splitting between Y$_{1}$ and Y$_{2}$. All values in units of cm$^{-1}$.}
\begin{tabular}{c c d d d d}
\hline
\hline
& & \multicolumn{2}{c}{Site 1} & \multicolumn{2}{c}{Site 2}\\
\hline
Multiplet & Energy & \textrm{CF} & \textrm{Fit} & \textrm{CF} & \textrm{Fit}\\
          &level   &    &              & & \\
\hline
\multirow{2}{*}{$^{5}$I$_{8}$}  & Z$_{1}$ & 0 & 0 & 0 & 0\\
 & Z$_{2}$ & 4.67 & 4.87 & 1.77 & 4.72\\
 \\
\multirow{2}{*}{$^{5}$I$_{7}$} & Y$_{1}$ & 5162.19 & 5158.20 & 5125.37 & 5135.25\\
 & Y$_{2}$ & 5162.70 & 5159.25 & 5125.92 & 5136.40\\
 \\
 & $\Delta_{Y_{1}Y_{2}}$ & 0.51 & 1.05 & 0.55 & 1.15\\
 \hline
 \hline
 \end{tabular}
\end{table}

Figure~\ref{fig:Site_2_0_0.3T_Zeeman} shows the lowest-energy $^5$I$_8$ (Z) $\longrightarrow$ $^5$I$_7$ (Y) absorption transitions, which we assign to Site 2 (see below). Both the ground and excited states are closely spaced electronic singlets, which can be observed as four sets of transitions in Fig.~\ref{fig:Site_2_0_0.3T_Zeeman}(a). At low temperatures the population of Z$_2$ is significantly less than that of Z$_1$.  Though the sample is in contact with a mount that is nominally at 4.2\,K, a comparison of intensity ratios with the Boltzmann equation suggests that the sample temperature is close to 10\,K. The close proximity of these levels, together with the low point-group symmetry of the Ho$^{3+}$ ions, suggests we may treat these levels as a pair of pseudo-doublets, with their close proximity giving significant magnetic interactions such as Zeeman-hyperfine effects. Indeed, the zero-field spectrum shown in Fig.~\ref{fig:Site_2_0_0.3T_Zeeman}(a) exhibits partially resolved hyperfine splittings significantly larger than the inhomogeneous broadening of the spectral lines. Estimated transition energies are listed in Table~\ref{tab:Site_2_0T_Para}.  Figure~\ref{fig:Site_2_0_0.3T_Zeeman}(b) shows the same transitions with the application of a 0.3 T magnetic field along the $b$ axis.

We model the spectrum by first calculating electronic energy levels and wavefunctions  and then treating the hyperfine interaction as a perturbation. 
Table~\ref{tab:Fitparameters} gives calculated energies for the lowest two $^5$I$_8$ (Z$_1$, Z$_2$) and  $^5$I$_7$ (Y$_1$, Y$_2$) states. These calculations use the crystal-field parameters for Er$^{3+}$ in Y$_2$SiO$_5$ from Ref.\ \cite{Horvath2016} for Site 1 and Site 2, and the free-ion parameters for Ho$^{3+}$ from Ref.\ \cite{carnall_systematic_1989}. The Y$_1$ and  Y$_2$ levels for Site 2  are calculated to be lower than those for Site 1, so we assign the spectrum in Fig.~\ref{fig:Site_2_0_0.3T_Zeeman} to Site 2. The calculated splitting between Z$_1$ and Z$_2$ and the splitting between  Y$_1$ and Y$_2$ is smaller than the experimental splitting, so we treat the energies as adjustable parameters.

Table~\ref{tab:Joperator} gives the relevant matrix elements of the  angular momentum operators ${J}_{x}$, ${J}_{y}$ and ${J}_{z}$ calculated using  eigenvectors from the crystal-field calculation. These operators contribute to both the hyperfine splitting (Eq.~\ref{eqn:AdotI}) and the magnetic splitting (Eq.~\ref{eqn:Zeeman}). The different magnitudes of the matrix elements of these operator for the two sites, and the different directions, will allow us to confirm our assignments. 
The $A$ and $g$ factors for the relevant multiplets are  given in  Table~\ref{tab:MagHFPara}. The only free parameters in our calculation are the energies of Z$_1$, Z$_2$, Y$_1$, and Y$_2$. These fitted energies are given in Table~\ref{tab:Fitparameters}.

\begin{table}[tb!]
\centering
\caption{\label{tab:Joperator}
Absolute values of matrix elements of the angular momentum operators from crystal-field calculations. The diagonal matrix elements are not shown, but are of the order of 10$^{-4}$.}
    \begin{tabular}{c c c c c c c}
     \hline
     \hline
    {} & \multicolumn{3}{c}{Site 1} & \multicolumn{3}{c}{Site 2} \\
     \hline
     Energy  & ${J}_{x}$ & ${J}_{y}$ & ${J}_{z}$ & ${J}_{x}$ & ${J}_{y}$ & ${J}_{z}$ \\
     level\\
     \hline
    $|\langle Z_1|{J}_{i} |Z_2\rangle|$ & 2.53 & 5.34 & 1.47 &3.01 & 2.75 & 5.12 \\
    \\
    $|\langle Y_1|{J}_{i} |Y_2\rangle|$ & 1.96 & 5.19 & 2.15 & 3.01 & 2.76 & 4.9924 \\
     \hline
     \hline
    \end{tabular}
\end{table}

\begin{table}[tb!]
\centering
 \caption{\label{tab:MagHFPara}
Hyperfine and magnetic constants calculated using eigenvectors of the free-ion part of the Hamiltonian \cite{Wells2004}.}
 \begin{tabular}{c c}
 \hline
 \hline
  Constant & Value \\
 \hline
 A($^{5}$I$_{8}$) & 812 MHz\\
 A($^{5}$I$_{7}$) & 883 MHz\\
 g($^{5}$I$_{8}$) & 1.24\\
 g($^{5}$I$_{7}$) & 1.17\\
 \hline
 \hline 
 \end{tabular}
\end{table}

\begin{figure}[tb!]
\centering
\includegraphics[width=\figwidthone\columnwidth]{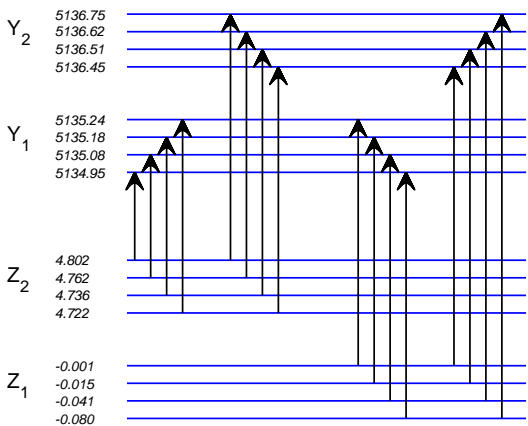}
\caption{\label{fig:HF_schematic}
Calculated hyperfine energy level structure for Site 2 of  Ho$^{3+}$:Y$_{2}$SiO$_{5}$. Vertical arrows show the strong transitions between Z$_{1}$-Z$_{2}$ and Y$_{1}$-Y$_{2}$.} 
\end{figure}

\begin{figure}[tb!]
\centering
\includegraphics[width=\figwidthfour\columnwidth]{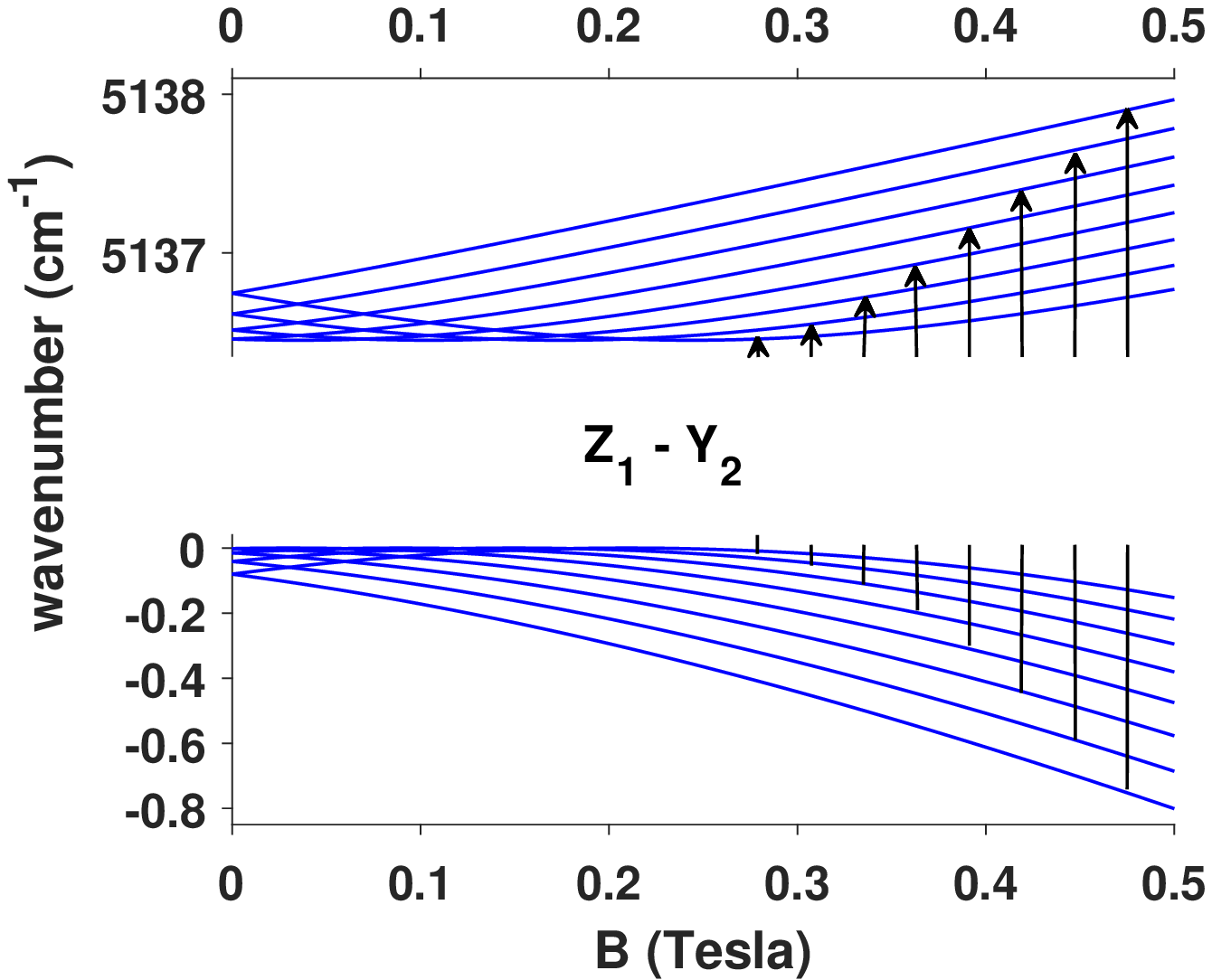}
\\
\bigskip
\includegraphics[width=\figwidthfour\columnwidth]{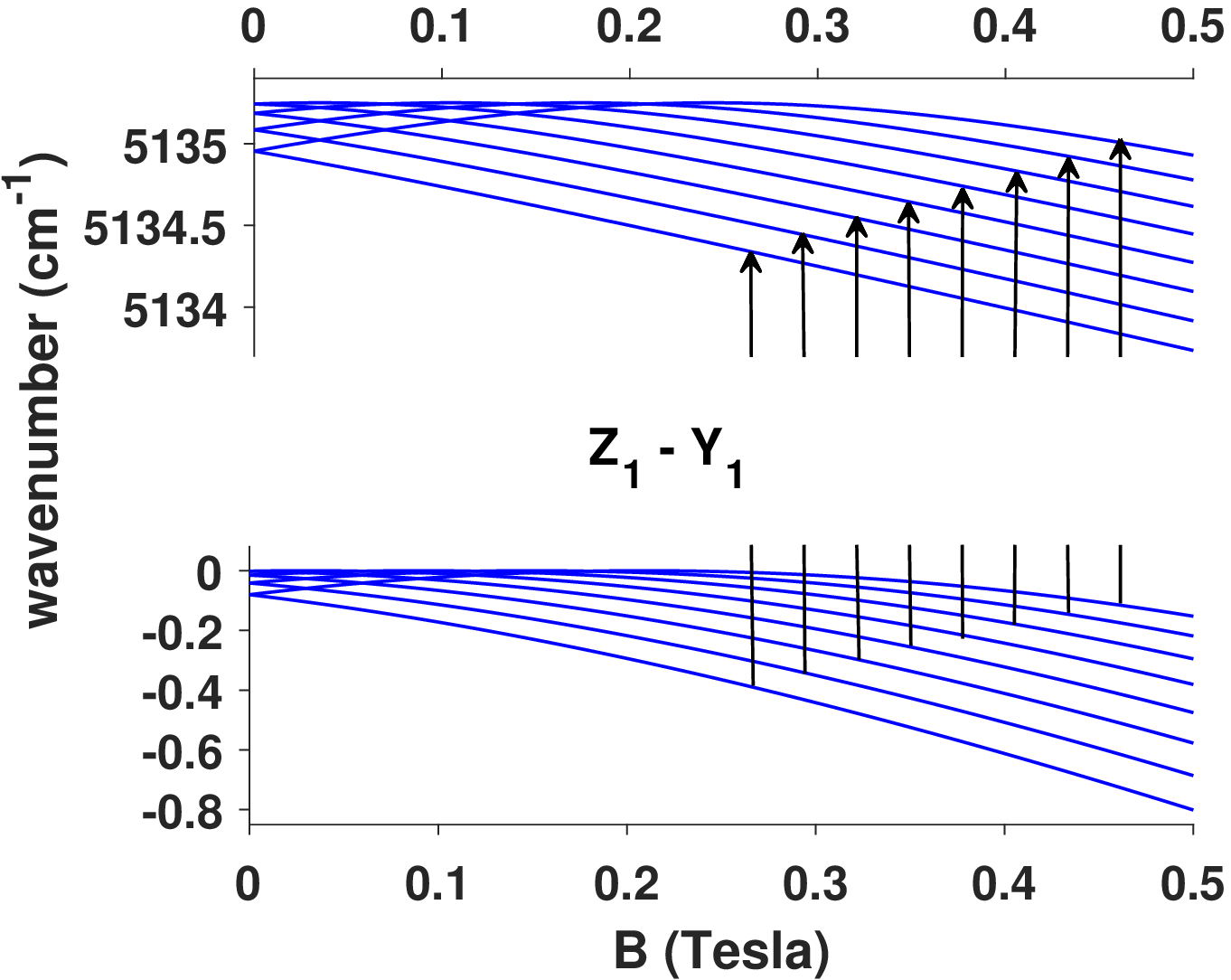}
\\
\bigskip
\includegraphics[width=\figwidthfour\columnwidth]{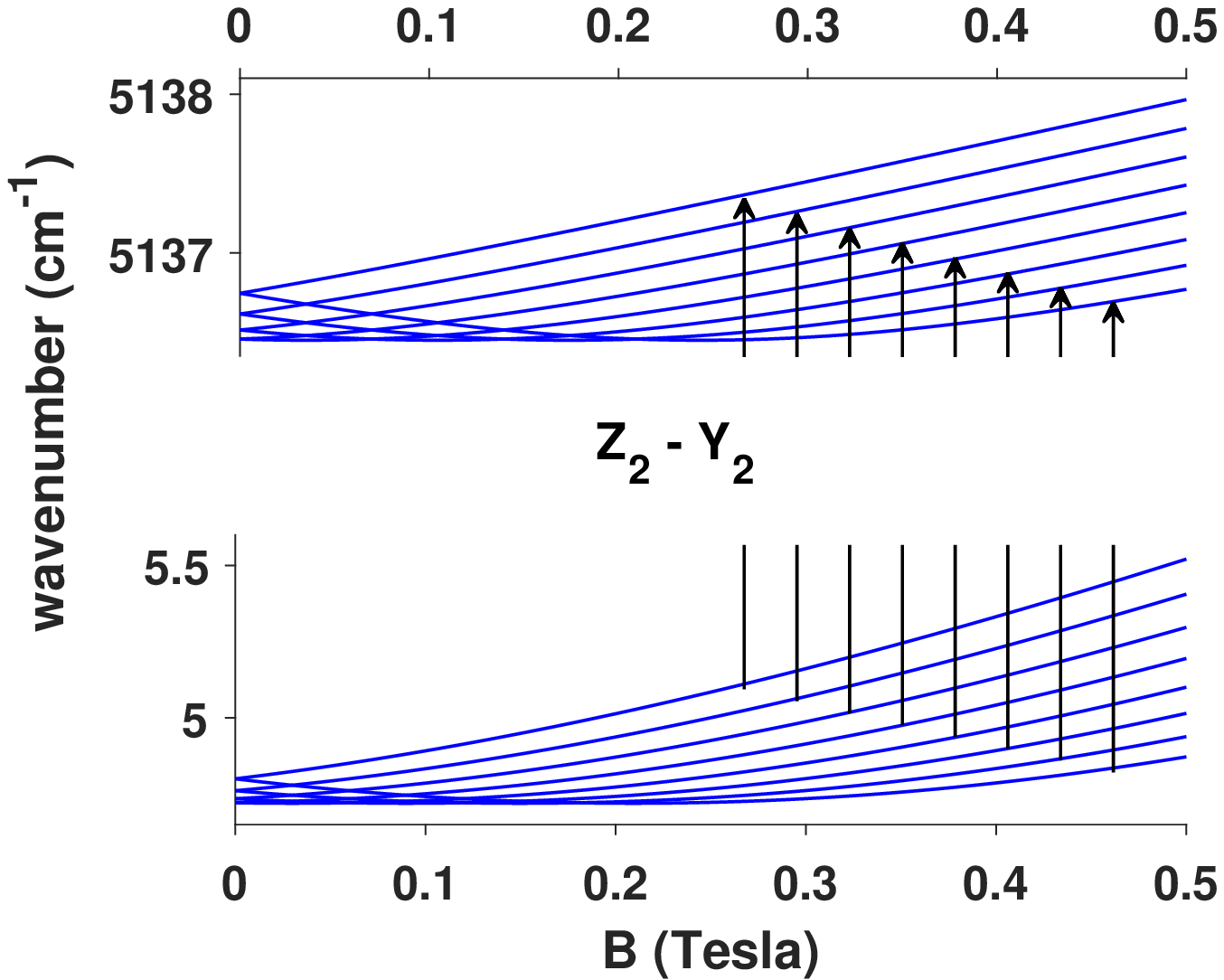}
\\
\bigskip
\includegraphics[width=\figwidthfour\columnwidth]{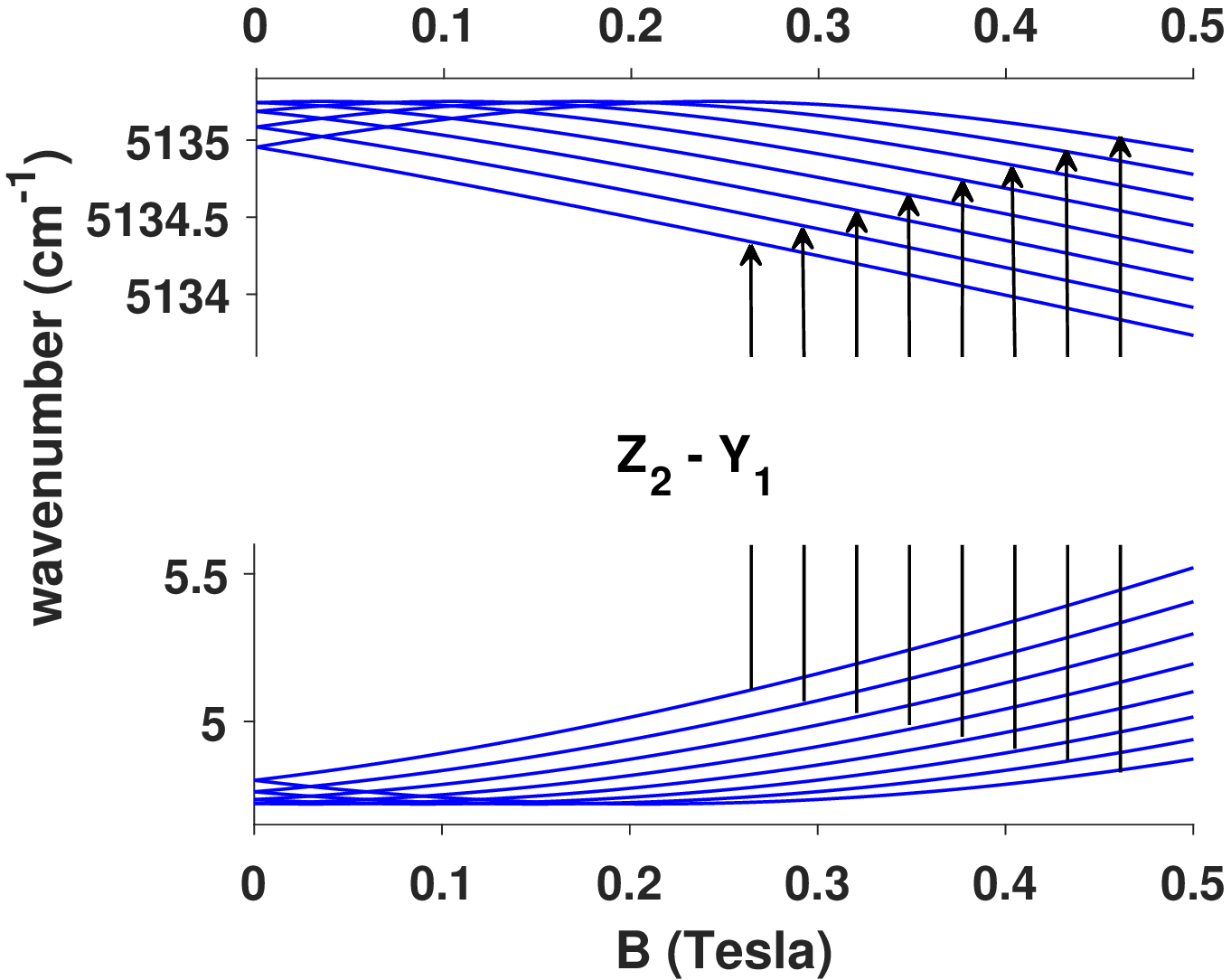}
\caption{\label{fig:Zeeman_schematic}
Calculated Zeeman-hyperfine energy level structure for Site 2 of  Ho$^{3+}$:Y$_{2}$SiO$_{5}$ with a magnetic field applied along the $b$ axis. Vertical arrows show the strong transitions between Z$_{1}$-Z$_{2}$ and Y$_{1}$-Y$_{2}$.}
\end{figure}

The addition of the nuclear spin ($I=7/2$) expands each singlet electronic state into 8 electronic-nuclear states. However, due to Kramers degeneracy, in the absence of a magnetic field there are only four distinct energies for each electronic state. Thus, there are two $16\times16$ matrices, one each for the Z$_1$-Z$_2$ and Y$_1$-Y$_2$ states. Since the diagonal matrix elements of $\mathbf{J}$ are extremely small (they would be zero for isolated singlets), it is the off-diagonal matrix elements of ${A_J}\mathbf{J}\cdot\mathbf{I}$  between Z$_1$ and Z$_2$ or Y$_1$ and Y$_2$ that are responsible for the hyperfine splitting. 
Figure~\ref{fig:HF_schematic} shows the calculated zero-field energy levels and the calculated transition energies are given in Table~\ref{tab:Site_2_0T_Para}. Figure~\ref{fig:Zeeman_schematic} shows the energies as a function of magnetic field.

The eigenvectors from the electronic-hyperfine calculation can also be used to estimate the absorption. The $^5$I$_8$  to $^5$I$_7$ transitions are predominantly magnetic dipole \cite{Wells2004} in nature. The calculated spectra in Fig.~\ref{fig:Site_2_0_0.3T_Zeeman} were obtained by calculating the squares of the appropriate magnetic-dipole moments, and adding line profiles of width (FWHM) 0.12\,cm$^{-1}$. 

In the low symmetry sites of Y$_2$SiO$_5$, the hyperfine Hamiltonian (Eq.~\ref{eqn:AdotI}) mixes states with different $M_I$. This is in contrast to the high-symmetry sites in Refs.~\cite{Wells2004,Boldyrev2019}, where the states are pure  $M_I$.  Since optical transitions do not change the nuclear spin, in the high-symmetry systems the number of allowed transitions is restricted, and at zero field there are only four distinct transitions  between singlet electronic states. Our spectra, and our calculations, also give only four strong transitions at zero field, even though in principle all possible transitions are allowed. This is because the eigenvectors of the hyperfine Hamiltonian for Z$_1$-Z$_2$ and Y$_1$-Y$_2$ are sufficiently similar that only four transitions with distinct energies at zero field, and 8 at non-zero field, have significant intensity. 

\begin{figure}[tb!]
\centering
\includegraphics[width=\figwidththree\columnwidth]{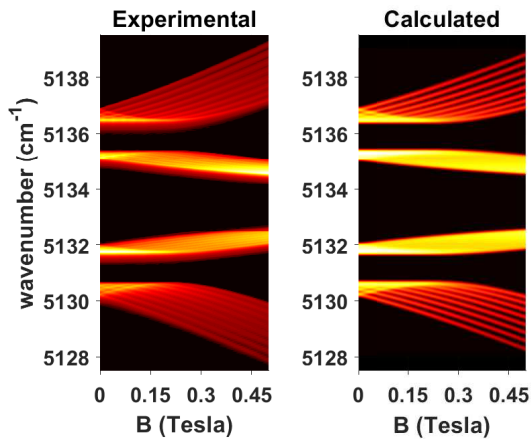}
\caption{\label{fig:Site_2_Zeeman}
Map of the experimental and calculated Zeeman-hyperfine infrared absorption spectra for the Z$_{1}$-Z$_{2}\longrightarrow$Y$_{1}$-Y$_{2}$ transitions of Site 2 for Ho$^{3+}$:Y$_{2}$SiO$_{5}$ with a magnetic field applied along the $b$ axis.}
\end{figure}

\begin{figure}[tb!]
\centering
\includegraphics[width=\figwidthone\columnwidth]{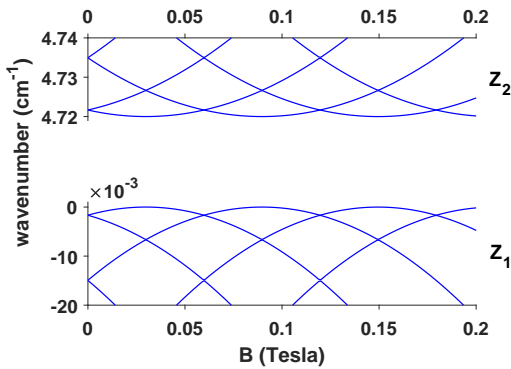}
\caption{\label{fig:AC}
Calculated anti-crossings in the Z$_{1}$-Z$_{2}$ region of Site 2 with a magnetic field applied along the $b$ axis. The energy gaps at the anti-crossing points \emph{within}  Z$_{1}$ and Z$_{2}$ are the order of 10$^{-4}$\,cm$^{-1}$ (5\.MHz), so are not visible in the Figure.}
\end{figure}

\begin{figure}[tb!]
\centering
\includegraphics[width=\figwidthtwo\columnwidth]{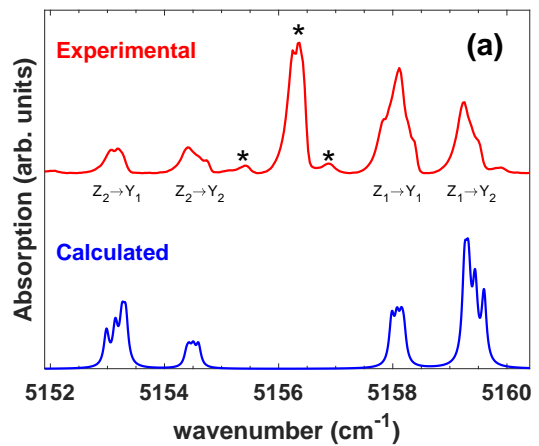}\\
\bigskip
\includegraphics[width=\figwidthtwo\columnwidth]{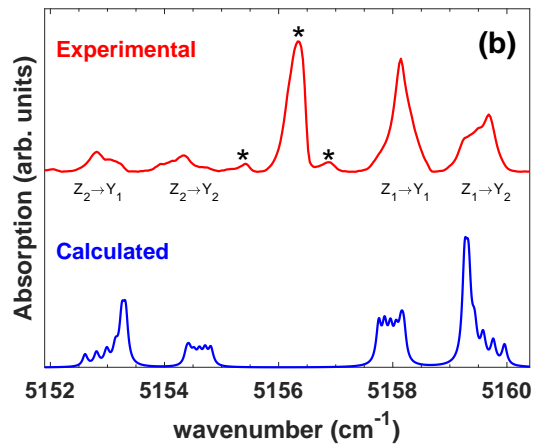}
\caption{\label{fig:Site_1_0_0_3T_Zeeman}
4.2~K absorption spectrum of the Z$_{1}$-Z$_{2}\longrightarrow$Y$_{1}$-Y$_{2}$ transitions for Site 1 of Ho$^{3+}$:Y$_{2}$SiO$_{5}$. (a) Zero field spectrum and (b) 0.3~T spectrum with a magnetic field applied along the $b$ axis.
Asterisks indicate transitions assigned to Site 2 (the largest feature) and water.}
\end{figure}

\begin{figure}[tb!]
\centering
\includegraphics[width=\figwidththree\columnwidth]{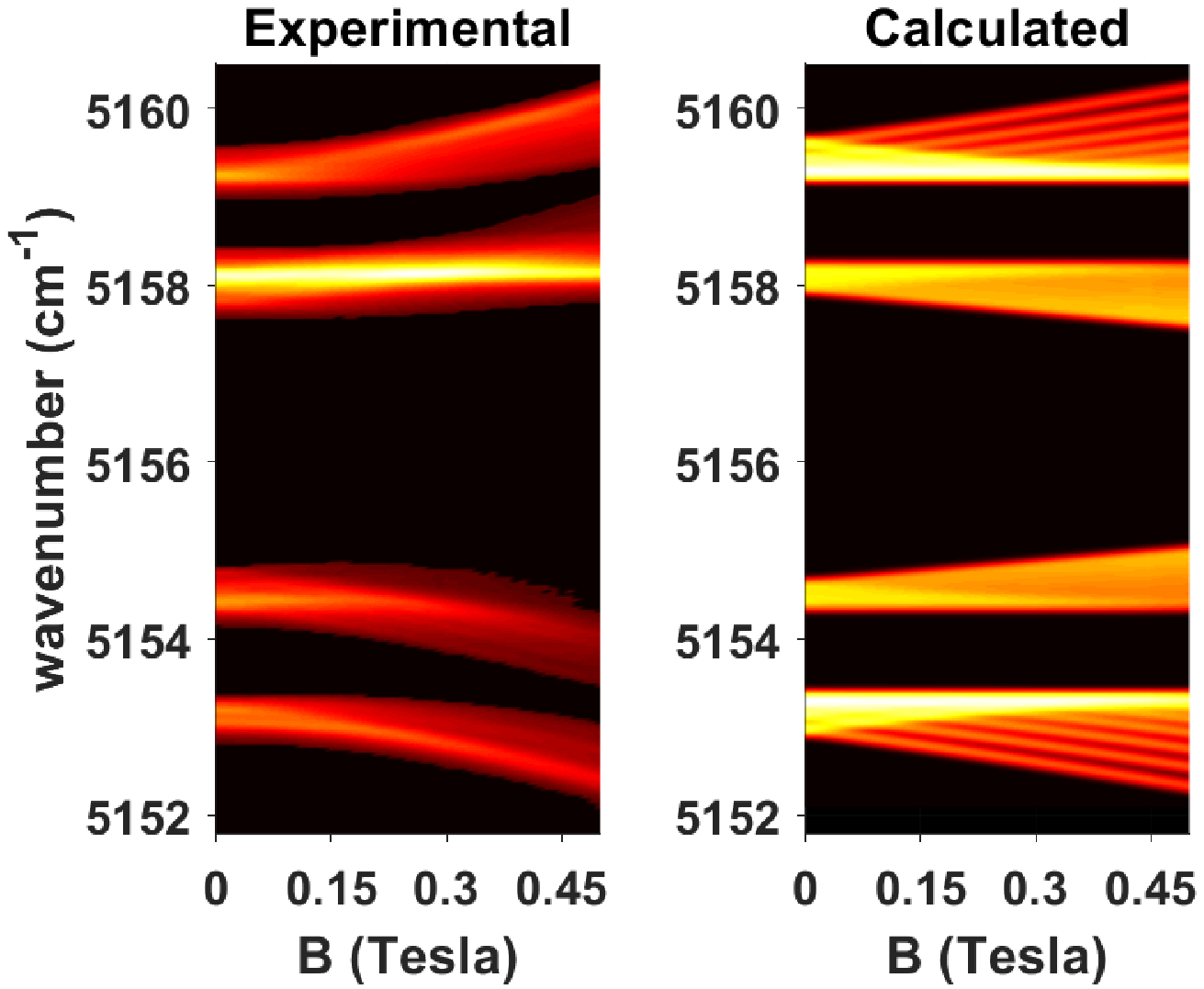}
\caption{\label{fig:Site_1_Zeeman}
Map of the experimental and calculated Zeeman-hyperfine infrared absorption spectra for the Z$_{1}$-Z$_{2}\longrightarrow$Y$_{1}$-Y$_{2}$ transitions of Site 1 for Ho$^{3+}$:Y$_{2}$SiO$_{5}$ with a magnetic field applied along the $b$ axis. For clarity, transitions assigned to Site 2 and water have been deleted from the experimental map. }
\end{figure}

Experimental data and calculations for the magnetic field range from 0 to 0.5\,T are shown in Fig.~\ref{fig:Site_2_Zeeman}. With the aid of Fig.~\ref{fig:Zeeman_schematic}, which indicates the approximate selection rules, it is clear why the Z$_1 \longrightarrow$Y$_2$ and  Z$_2 \longrightarrow$Y$_1$ transitions move apart, and the Z$_1 \longrightarrow$Y$_1$ and  Z$_2 \longrightarrow$Y$_2$ transitions move together as the field increases. In the latter case the energy \emph{differences} tend towards the same value. A similar calculation also reproduces the magnetic splitting along the D$_{2}$ axis. The matrix elements of $J_y$ are smaller than those of $J_z$ (Table~\ref{tab:Joperator}), so both experimental and calculated magnetic splittings are smaller. Note, also, that the  $J_z$ matrix elements are significantly smaller for Site 1, and if the Site 1 eigenvectors were used, the calculation would be in poor agreement with the experimental measurements. Thus, the calculations confirm our site assignment. 

Anti-crossings between magnetic-hyperfine levels are of interest in quantum-information applications, since transitions at those points suffer minimal effects of magnetic fluctuations \cite{,zhong_optically_2015,Shiddiq2016}. 
Figure~\ref{fig:AC} gives a blowup of our calculated Z$_1$ and Z$_2$ states. As the magnetic field is increased,  anticrossings between the Z$_1$ and Z$_2$ states are apparent. These are the order of 4.7\,cm$^{-1}$ (140\,GHz). This splitting is much larger than the 9\,GHz splitting utilized in the EPR experiments of Reference \cite{Shiddiq2016}, due to the much larger matrix elements of $\mathbf{J}$ connecting the Z$_1$ and Z$_2$ states in our system. Consequently,  the second derivatives with respect to changes in the magnetic field are considerably smaller in our system. 

There are also anti-crossings between the  Z$_1$ states. These are not resolved in Fig.~\ref{fig:AC}, since the spacings at the anti-crossings range from 2 to 9\,MHz. These spacings are small because the  matrix elements of  $\mathbf{J}$ within  Z$_1$ are the order of $10^{-4}$\,cm$^{-1}$, which is a result of the electronic states being non-degenerate.  In Reference \cite{Boldyrev2019}, anticrossings of approximately 1.8\,GHz were  observed by infra-red absorption in a degenerate excited state of Ho$^{3+}$ in LiYF$_4$. The hyperfine anticrossings in the non-degenerate electronic states of Ho$^{3+}$:Y$_{2}$SiO$_{5}$ are too small to resolve by optical absorption measurements. 

We now consider a region that we assign to transitions of Ho$^{3+}$ in Site 1. In Figs.~\ref{fig:Site_1_0_0_3T_Zeeman} and \ref{fig:Site_1_Zeeman}
experimental data is compared with calculations of hyperfine and magnetic splittings for Z$_{1}$-Z$_{2}\longrightarrow$Y$_{1}$-Y$_{2}$ transitions of Site 1. This region also contains lines that we assign to Site 2 and to water (indicated by asterisks). For clarity, these features were removed from Fig.~\ref{fig:Site_1_Zeeman}. The hyperfine structure is not fully resolved, but it is clear that both measured and calculated Zeeman splittings are much smaller for Site 1 than for Site 2. Though the calculation reproduces the experimental \emph{energies}, it does not accurately reproduce the experimental \emph{intensities} under the influence of a magnetic field  (Fig.~\ref{fig:Site_1_0_0_3T_Zeeman}(b)). It is possible that mixing with higher-lying electronic states is responsible for this discrepancy. To properly acccount for such effects would require going beyond the perturbation approach used here, to a  calculation taking into account the crystal-field, hyperfine, and Zeeman interactions for the full 4$f^{10}$ configuration.

\section{Conclusions}

We have presented Zeeman-hyperfine spectra for the lowest energy infrared absorption transitions of Ho$^{3+}$ doped into high-quality Y$_{2}$SiO$_{5}$ crystals. We have demonstrated that the data may be modeled  using electron-nuclear wavefunctions derived from a Ho$^{3+}$ crystal-field calculation using parameters obtained previously for the Er$^{3+}$ ion, which neighbors Ho$^{3+}$ in the periodic table. The results presented here form an important test of both the transferability of parameters from ion-to-ion and the predictive ability of crystal-field calculations for sites exhibiting very low $C_{1}$ point group symmetry. The calculations are, therefore,  an important step in establishing a consistent set of crystal-field parameters for rare-earth ions in Y$_{2}$SiO$_{5}$. It is notable that not only are our calculations  able to distinguish spectroscopic features of the two substitutional sites but in fact, give excellent agreement with few fitting parameters. 

\section*{Acknowledgments}
SM acknowledges support from the University of Canterbury in the form of University Doctoral Scholarship. 
EL-H acknowledges support from DGA. 
The technical assistance of Mr Stephen Hemmingson, Mr Robert Thirkettle and Mr Graeme MacDonald (UoC, NZ) is gratefully acknowledged.


%

\end{document}